\documentclass[aps,showpacs,nofootinbib]{revtex4}

\usepackage{graphicx}

\newcommand{\be}{\begin{equation}}
\newcommand{\ee}{\end{equation}}
\newcommand{\ben}{\begin{eqnarray}}
\newcommand{\een}{\end{eqnarray}}

\newcommand{\la}{{\lambda}}

\newcommand{\p}{\partial}
\newcommand{\na}{\nabla}

\newcommand{\Lie}{{\cal L}}

\newcommand{\ga}{\gamma}


\begin{document}

\title{Uniqueness theorem for stationary black ring solution of $\sigma$-models in
five dimensions}
%%%%%%%%%%%%%%%%%%%%%%%%%%%%%%%%%%%%%%%%%%%%%%%%%%%%%%%%%%%%%%
%\author{David Langlois}
%\affiliation{Institute of  \protect \\
%University \protect \\
%20-031 , France \protect \\
%david@tytan.umcs.lublin.fr \protect \\

%\author{ Rafa{\l} Moderski}
%\address{
%Nicolaus Copernicus Astronomical Center \protect \\
%Polish Academy of Sciences \protect \\
%00-716 Warsaw, Bartycka 18, Poland \protect \\
%moderski@camk.edu.pl }

\author{Marek Rogatko}
\affiliation{Institute of Physics \protect \\
Maria Curie-Sklodowska University \protect \\
20-031 Lublin, pl.~Marii Curie-Sklodowskiej 1, Poland \protect \\
and \protect \\
Institute des Hautes Etudes Scientifiques \protect \\
35 route de Chartres, 91440 Burres-sur-Yvette, France \protect \\
rogat@tytan.umcs.lublin.pl \protect \\
rogat@kft.umcs.lublin.pl}

%%%%%%%%%%%%%%%%%%%%%%%%%%%%%%%%%%%%%%%%%%%%%%%%%%%%%%%%%%%%%%%%%%%%
\date{\today}
%\pacs{04.30.Nk, 04.40.-b}

%%%%%%%%%%%%%%%%%%%%%%%%%%%%%%%%%%%%%%%%%%%%%%%%%%%%%%%%%%%%%%%%%%%%%%
\begin{abstract}
We study stationary axisymmetric self-gravitating non-linear $\sigma$-model
in five-dimensional spacetime admitting three commutating Killing vector fields.
We show that the only asymptotically flat black ring solution with a regular 
rotating event horizon is the black ring characterized by mass and two angular momenta
with constant mapping.

\end{abstract}
%%%%%%%%%%%%%%%%%%%%%%%%%%%%%%%%%%%%%%%%%%%%%%%%%%%%%%%%%%%%%%%%%%%%%%%

\maketitle

%%%%%%%%%%%%%%%%%%%%%%%%%%%%%%%%%%%%%%%%%%%%%%%%%%%%%%%%%%%%%%%%%%%%%%%
\section{Introduction}
It has been renewed of interests in higher dimensional black hole solutions
trigerred by the unification attempts, both in the context of M/string theory as
well as the brane world scenario.
The unification attempts such as M/string theory
described our Universe
as a brane or defect emerged in higher dimensional geometry. 
$E8 \times E8$ heterotic string theory
at strong coupling may be described in terms of M-theory acting in eleven-dimensional
spacetime with boundaries where ten-dimensional Yang-Mills gauge theories reside on two
boundaries \cite{hor}. On the other hand,
the so-called
TeV gravity acquires much attention to higher dimensional black hole 
because of the idea that such kind of objects may be 
produced in the near future, in high energy experiments \cite{gid}.
This kind of black holes (the so-called mini black holes) are classical solutions of higher dimensional 
Einstein's equations. The radius of their event horizon is much smaller
that the scale of extra dimensions.
\par
In four-dimensional spacetime the classification of non-singular black hole solutions
began with the Israel's work
\cite{isr}, then
M\"uller zum Hagen {\it et al.} \cite{mil73} and Robinson \cite{rob77},
provided other
contributions to the problem \cite{bun87,ru,ma1,he1,he93}.
Both for vacuum and Einstein-Maxwell (EM) black holes
the condition of non-degeneracy of the 
event horizon was removed Refs.\cite{chr99a,chr99b}. It was proved that for the static
electro-vacuum black holes all degenerate components of 
the event horizon should have charges of the same signs.
However, recently this last restriction was removed from the theorem \cite{chr06,chr07}.
\par
The problem of uniqueness theorem for stationary axisymmetric black hole
was considered in Refs.\cite{car,rob} and the complete proof was provided by
Mazur \cite{maz} and Bunting \cite{bun}
(see also for a review of the uniqueness of black hole
solutions story see \cite{book} and references therein).
\par
Studies of the low-energy string theory 
also renew the 
resurgence of works concerning the mathematical aspects of the
black holes emerging in it. Namely,
the staticity theorem for Einstein-Maxwell axion dilaton (EMAD) gravity
was studied in Ref.\cite{sta} and
uniqueness of the black hole solutions in dilaton gravity was  proved in 
works
\cite{dil,mar01}, while the uniqueness of the static
dilaton $U(1)^2$
black holes being the solution of $N = 4, d = 4$ supergravity
was provided in \cite{rog99}. The extension of the proof
to
$U(1)^{N}$ static dilaton black holes was established in Ref.\cite{rog02}.
\par
The possibility of production of higher dimensional black holes in 
accelerators caused 
the considerable interests in
$n$-dimensional black hole uniqueness theorem, both in vacuum and
charged case \cite{gib03,gib02a,gib02b}. 
The uniqueness theorem for self-gravitating nonlinear $\sigma$-models
in higher dimensional spacetime was obtained in \cite{rog02a}.\\
The complete 
classification of $n$-dimensional
charged black holes having both degenerate and non-degenerate
components of event horizon was initiated in Ref.\cite{rog03}.
Studies of the near-horizon geometry of degenerate horizons enable to eliminate the previuos
restriction bounded with the inequality satysfied by the charges of the adequate components of
the aforementioned black hole horizons \cite{rog06}. 
Uniqueness theorem for $n$-dimensional static black hole carrying
{\it electric} and {\it magnetic} components of $(n-2)$-gauge
containing an asymptotically flat hypersurface with compact interior and nondegenerate 
components of the event horizon was presented
in Ref.\cite{rog04}.
The staticity theorem for generalized Einstein-Maxwell (EM) system in $n$-dimensional
spacetime was presented in \cite{rog05}.
\par
It was argued that future experiments of colliding high energy particles
held great promise for iluminating the nature of mini black holes.
One of the first
signature of appearing black holes in future
accelerator experiments will be a Hawking emission and generically
the black holes will have angular momentum. This fact causes that the uniqueness
of rotating black holes in higher dimensions is of a great importance in
studies mathematical properties of them.
A salient feature of the axisymmetric stationary solutions in higher dimensions is
the fact that they can admit event horizon with nonspherical topology
in contrast to four-dimensional case. The topology of the event horizon can not be
uniquely determined. For example in five-dimensional case one has the topology of
$S^{3}$ sphere or $S^{1} \times S^{2}$ \cite{cai01}, while in higher dimensions 
the topology is far more complicated \cite{hel06,gal06}. It was advocated that generalization 
of the Kerr metric to arbitrary $n$-dimensions proposed by Myers and Perry \cite{mye86}
is not unique. The counterexample showing that a five-dimensional
rotating black hole ring solution has the same angular momentum and mass
as five-dimensional axisymmetric stationary black hole. But as far as topology was concerned,
its event horizon was homeomorphic to $S^{1} \times S^{2}$ 
\cite{emp02,emp04}). In Ref.\cite{mor04} it was shown that Myers-Perry
solution is the unique black hole in five-dimensions in the class of spherical topology
with three commuting Killing vectors.
On the other hand, the uniqueness for stationary self-gravitating nonlinear $\sigma$-models
in five-dimensional spacetime was obtained in \cite{rog04a}, where it was shown that vacuum
Myers-Perry Kerr solution is the only maximally extended stationary axisymmetric asymptotic 
flat solution having the regular event horizon with constant mapping.
\par
It was proved in Ref.\cite{hol07} that a higher dimensional stationary rotating black 
hole must be axisymmetric with no assumptions concerning the topology of the horizon cross-section
other than compactness of it. The also assume that the horizon is non-degenerate and spacetime is analytic.\\
In Ref.\cite{hol07a} the authors showed that two asymptotically flat five-dimensional black hole solutions
of vacuum Einstein Eqs. with the same topology, mass and angular momentum and moreover with the same {\it rod structure} \cite{har}
are isometric to each other. Then, the proof was generalized to include Maxwell fields \cite{hol07b}.
\par
Recently, it was shown in Ref.\cite{mor08} that assuming the existence of two additional 
commutating axial Killing vector fields and the horizon topology of black ring $S^1 \times S^2$,
the only asymptotically flat black ring solution with a regular horizon is the Pomeransky-Sen'kov (PS)
black ring \cite{pom06}.
\par
Finding the black ring solutions triggered interests in these blossoming subject. Many other black ring,
or one should say {\it black object} solutions were found e.g., {\it black Saturn} (an object consisting of
rotating black rings with a spinning black hole as its center), the di-ring, the bi-ring etc. The recent
summary and review of our understanding of the aforementioned problem is presented in Ref.\cite{emp08}. 
Also thermodynamics of these black objects were intessively studied (see e.g., Refs.\cite{ther}).
\par
In our paper we shall treat the problem of uniqueness of five-dimensional
axisymmetric, stationary black ring solution for self-gravitating non-linear $\sigma$-models. 
In Sec.II we present general self-gravitating $\sigma$-model and
establish  the main result of
our work that the only regular black ring solution with regular rotating event horizon
is the five-dimensional vacuum PS black ring with constant mapping.

%%%%%%%%%%%%%%%%%%%%%%%%%%%%%%%%%%%%%%%%%%%%%%%%%%%%%%%%%%%%%%%%%%%%%%%%%%%%%%%%%%%%%%%%%%
\section{Five-dimensional rotating $\sigma$-models}
In this section we shall derive equations of motion for $n$-dimensional self-gravitating $\sigma$-model
being subject to the following action:
\be
I = \int d^{n}x~\sqrt{- {}^{(n)} g} \bigg[
{}^{(n)}R - {1\over 2} G_{AB}(\varphi (x))~ \varphi^{A}_{, \mu}
\varphi^{B ,\mu} \bigg].
\ee
The equations of motion for our non-linear $\sigma$-model can be derived from the variational principle.
They yield
\ben
\na_{\ga}\na^{\ga} \varphi^{A} &+& \Gamma^{A}_{BC}~
\varphi^{B}_{,\mu} \varphi^{C ,\mu} = 0,\\
{}^{(n)}G_{\mu \nu} &=& T_{\mu \nu}(\varphi),
\een
where the energy momentum for the underlying model has the form  
\be
T_{\mu \nu}(\varphi) = G_{AB}(\varphi (x))~ \varphi^{A}_{,\mu} 
\varphi^{B}_{,\nu} - {1 \over 2} G_{AB}(\varphi (x))~ 
\varphi^{A}_{,\ga} \varphi^{B ,\ga} g_{\mu \nu}.
\ee
In what follows we shall take into account the asymptotically, five-dimensional
flat spacetime, i.e., the spacetime will contain a data set 
$(\Sigma_{end}, g_{ij}, K_{ij})$ with scalar fields of $\varphi$
such that a spacelike hypersurface $\Sigma_{end}$ is diffeomorphic to $\bf R^{4}$ minus 
a ball. The asymptotical conditions of the following forms should also be satisfied:
\be
\mid g_{ij} - \delta_{ij} \mid +
r \mid \p_{a} g_{ij} \mid + \dots
+r^{m} \mid \p_{a_{1} \dots a_{m}} g_{ij} \mid + r \mid K_{ij}\mid + \dots
+r^{m} \mid \p_{a_{1} \dots a_{m-1}} K_{ij} \mid \le {\cal O}\bigg( {1 \over r} \bigg),
\ee
where $g_{ij}$ and $K_{ij}$ are induced on $\Sigma_{end}$. $K_{ij}$ is the extrinsic
curvature tensor of the hypersurface $\Sigma_{end}$. 
It is required
that in the local coordinates on $\Sigma_{end}$ the scalar
field satisfies the following fall-off condition:
\be
\varphi^{A} =
\varphi^{A}_{\infty} + {\cal O}\bigg( {1 \over r^{3/2}} \bigg).
\ee
As we shall consider statationary axisymmetric five-dimensional spacetime, thus it
will admit three commutating Killing vector 
fields $k_{\mu}, \phi_{\mu}, \psi_{\mu}$
\be
[k, \phi] = [k, \psi] = [\phi, \psi] = 0.
\ee
$k_{\mu}$ is an asymptotically timelike Killing vector field for which $V = - k_{\mu}k^{\mu}$, while
$\phi_{\mu}$ and $\psi_{\mu}$ are spacelike Killing vector fields. They all have closed orbits.
Then, denoting by $\Lie$ the Lie derivative with respect to the adequate Killing vector
fields one obtains the following:
\be
\Lie_{k}~g_{\mu \nu} = \Lie_{\phi}~g_{\mu \nu} = \Lie_{\psi}~g_{\mu \nu} = 0,
\ee
 The scalar field $\varphi$ will be also invariant due to 
the action of Killing vector fields. Namely, we have 
\be
\Lie_{k}~\varphi = \Lie_{\phi}~\varphi  = \Lie_{\psi}~\varphi = 0.
\ee
\par
The metric of the general black ring solution is given by \cite{mor08}
\be
ds^2 = {H(y, x) \over H(x, y)} \bigg( dt + \Omega \bigg)^2 - {F(x, y) \over H(y, x)} d\phi^2
- 2 {J(x, y) \over H(y, x)}d\phi~d\psi + {F(x, y) \over H(y, x)}d\psi^2 +
{2 k^2 H(x, y) \over (x - y)^2 (1 - \nu)^2} \bigg( {dx^2 \over G(x)}
- {dy^2 \over G(y)} \bigg),
\ee
where the range of $x, y$ coordinates is $- 1 \le x \le 1$ and 
$(- \la + \sqrt{\la^2 - 4 \nu})/2 \le y < \infty$ or
$- \infty < y \le -1$. The considered solution has four independent parameters 
which are subject to the inequalities. Namely, $0 \le \nu < 1$,~$2 \sqrt{\nu} \le
\la < 1 + \nu$,~$ k > 0$, and $ c \le b <1$ with the additional condition
of $c$ being equal to $\sqrt{ \la^2 - 4 \nu} / (1 - \nu)$.
It happenned that constant $c$ has the geometrical meaning as being the ratio
of the radius of $S^2$ to the radius $S^1$. One can introduce the canonical coordinates
$(\rho,~z)$ (which explicit form is given in Ref.\cite{mor08}). This enables us to write
the underlying metric
in the Weyl-Papapetrou form as 
\be
ds^2 = - {\rho^2 \over f} dt^2 + f_{ab} \bigg(
dx^{a} + \omega^{a} dt \bigg) \bigg(
dx^{b} + \omega^{b} dt \bigg) + {e^{2 \sigma} \over f}
\bigg( d\rho^2 + dz^2 \bigg),
\label{pap}
\ee
where all functions appearing in the above line element have the only $\rho$ and $z$ dependence.
Furthermore,
the metric (\ref{pap}) can be rearrange in the form which implies
\be
ds^2 = \sigma_{ab} dx^{a} dx^{b} + \ga_{ij} dx^{i} dx^{j},
\ee
where $a, b = t, \phi, \psi$ and comprises the first two components in expression (\ref{pap})
while  $i, j = \rho, z$ and describes $\ga_{ij} = X g_{ij}$. $g_{ij}$
stands for the metric of a flat spacetime written in $(\rho, z)$ coordinates.
The conformal factor is equal to 
$X = e^{2 \sigma}/f$.
Using rules of conformal transformation, after some algebra, we find expressions for the 
Ricci tensor components:
\ben
R_{ij} = {}^{(\ga)} R_{ij} &+& {1 \over 2} \ga_{ij} {}^{(\ga)} \na^2 \ln X -
{1 \over 2 \rho X} \bigg(
{}^{(\ga)} \na_{i} \rho~ {}^{(\ga)} \na_{j} X + {}^{(\ga)} \na_{i}X~ {}^{(\ga)} \na_{j} \rho
- \ga_{ij} {}^{(\ga)} \na^{k} \rho~ {}^{(\ga)} \na_{k} X
\bigg)  \\ \nonumber
&-& {1 \over \rho} {}^{(\ga)} \na_{i}{}^{(\ga)} \na_{j} \rho + {1 \over \rho^2}
{}^{(\ga)} \na_{i} \rho~ {}^{(\ga)} \na_{j} \rho +
{1 \over 4} {}^{(\ga)} \na_{i} \sigma^{ab}~ {}^{(\ga)} \na_{j} \sigma_{ab}.
\een
Consequently equations of motion may be written as
\ben
R_{\rho \rho} - R_{z z} &+& {1 \over \rho X} {}^{(g)}\na_{\rho} X - {1 \over \rho^2}
+ {1 \over 4} \bigg(
{}^{(g)}\na_{z} \sigma^{ab} {}^{(g)}\na_{z} \sigma_{ab} -
{}^{(g)}\na_{\rho} \sigma^{ab} {}^{(g)}\na_{\rho} \sigma_{ab}
\bigg) = {2 \over \rho}~ {}^{(g)} \na_{\rho} \sigma , \\
2 R_{z\rho} &+& {1 \over \rho X} {}^{(g)}\na_{z} X - 
{}^{(g)}\na_{z} \sigma^{ab} {}^{(g)}\na_{\rho} \sigma_{ab} = 
{2 \over \rho}~ {}^{(g)}\na_{z} \sigma, \\
R_{zz} + R_{\rho \rho} &-& {}^{(g)}\na_{m}{}^{(g)}\na^{m} \ln X +
{1 \over \rho X} {}^{(g)}\na_{\rho}X - {1 \over \rho X}
{}^{(g)}\na^{j} \rho {}^{(g)}\na_{j}X - {1 \over \rho^2} + \\ \nonumber
&-& {1 \over 4} \bigg(
{}^{(g)}\na_{\rho} \sigma^{ab}~{}^{(g)}\na_{\rho} \sigma_{ab} +
{}^{(g)}\na_{z} \sigma^{ab}~{}^{(g)}\na_{z} \sigma_{ab}
\bigg) = - 2 {}^{(g)}\na^{j} {}^{(g)}\na_{j} \sigma,
\een
where ${}^{(g)}\na$ is the derivative with respect to $g_{ij}$ metric.
The linearity of the above equations
implies directly that $\sigma(\rho, z)$ reduces to the sum of two
components as follows:
\be
\sigma = \sigma(vac) + \sigma(\varphi),
\label{ss}
\ee
where $\sigma(vac)$ is the solution of five dimensional vacuum equations of motion
while $\sigma(\varphi)$ is connected with the solution of matter equations.\\
On the other hand,
equations of motion for self-gravitating non-linear $\sigma$-model provide the following:
\be
{1 \over \rho} 
{}^{(g)}\na_{z} \sigma(\varphi) = {1\over 2} G_{AB}(\varphi (x)) 
\bigg( 
{}^{(g)}\na_{\rho}\varphi^{A}~ {}^{(g)}\na_{z}\varphi^{B} + 
{}^{(g)}\na_{z}\varphi^{A}~ {}^{(g)}\na_{\rho}
\varphi^{B} \bigg),
\label{m1}
\ee
\be
{1 \over \rho}
{}^{(g)}\na_{\rho} \sigma(\varphi) = {1\over 2} G_{AB}(\varphi (x)) 
\bigg( 
{}^{(g)}\na_{\rho} \varphi^{A}~ {}^{(g)}\na_{\rho} \varphi^{B} 
- {}^{(g)}\na_{z} \varphi^{A}~
{}^{(g)} \na_{z} \varphi^{B} \bigg),
\label{m2}
\ee
\be
{}^{(g)}\na_{m} {}^{(g)}\na^{m} \sigma (\varphi) = -{1\over 2} G_{AB}(\varphi (x)) 
\bigg( 
{}^{(g)}\na_{\rho} \varphi^{A} {}^{(g)}\na_{\rho} \varphi^{B}
+ {}^{(g)}\na_{z} \varphi^{A}
{}^{(g)}\na_{z} \varphi^{B} \bigg).
\label{m3}
\ee
In order to prove the uniqueness theorem for five-dimensional 
axisymmetric stationar non-linear
$\sigma$-model
we shall use the idea presented in Ref.\cite{heu95}. First, one chooses a
two-dimensional vector in the form as
\be
\Pi_{j} = \rho~ {}^{(g)}\na_{j} e^{- \sigma(\varphi)}.
\ee
Then, by virtue of
Stokes' theorem for $\Pi_{j}$ vector and integration over
the region $\Sigma = \{ (\rho,~z) \mid \rho \ge 0,~ - \infty < z < \infty \}$
we get
\ben \label{stok}
D_{\p {\Sigma}} &=& 
\int_{\p {\Sigma}} \rho e^{- \sigma(\varphi)}
\bigg( 
 {}^{(g)}\na_{z} 
\sigma(\varphi)~ d\rho -  {}^{(g)}\na_{\rho} \sigma(\varphi)~ dz \bigg)
\\ \nonumber
&=&
\int_{\Sigma} d\rho dz~ \rho e^{- \sigma(\varphi)}
\bigg[ {}^{(g)}\na^{i}  \sigma(\varphi) {}^{(g)}\na_{i} \sigma(\varphi)
- \bigg( {1 \over \rho} {}^{(g)}\na_{\rho}\sigma(\varphi) +
{}^{(g)}\na^{i} {}^{(g)}\na_{i}\sigma(\varphi) \bigg) \bigg].
\een 
From Eqs.(\ref{m1}),(\ref{m2}) and (\ref{m3}) 
it follows in particular that the second term on the right-hand 
side of (\ref{stok}) is greater or equal to zero. 
It implies that the right-hand side is the 
sum of two non-negative terms.\\
Now, let us calculate the left-hand side of expression (\ref{stok}).
In order to do so 
we have to decompose the integral over the segments of the rod and the integral over infinity.
So let us give a brief account of the {\it rod structure} of the underlying black ring \cite{har,mor08}. Namely,
the {\it rod structure} we should take into account is as follows:
\begin{enumerate}
\item{
the semi-infinite spacelike rod $[ - \infty,~ - ck^2 ]$ and the finite rod $[ck^2,~k^2]$ have the direction
$v = (0,~0,~1)$. It means that for $\rho = 0$,~$z \in [- \infty,~ - ck^2]$ and for $\rho = 0$ and
$z \in [ck^2,~k^2]$ the following relation is fulfilled $g_{ij} v^{j} = 0$. Because of the fact that
$g_{\psi \psi} = 0$ the conditions $\rho = 0$,~$z \in [- \infty,~ - ck^2]$ together with
$\rho = 0$ and
$z \in [ck^2,~k^2]$ denote $\psi$-axis,}
\item{
the finite timelike rod for the coordinates range $\rho = 0$ and
$z \in [- ck^2,~ck^2]$ with $g_{ij} v^{j} = 0$. It corresponds to the event horizon
with topology $S^1 \times S^2$. The Killing vector field $\psi_{\mu}$ vanishes on both sides of this
rod, while vector $v_{j}$ has values $(1,~ \Omega_{\phi},~\Omega_{\psi})$, where $\Omega_{\phi}$
denotes the angular velocity along the direction of $\phi_{\mu}$ Killing vector field and
$\Omega_{\psi}$ is the angular velocity along the direction of Killing vector $\psi_{\mu}$.}
\item{
The semi-infinite spacelike rod for the range of coordinates $\rho = 0$,~$z \in [k^2,~\infty]$, with the direction vector
$v = (0,~1,~0)$. Since one has that $g_{\phi \phi} = 0$ than it denotes $\phi$-axis.}
\end{enumerate}
Having in mind the {\it rod structure} of the black ring solution,
as we have mentioned, one has to decompose the boundary integral on the left-han side of Eq.(\ref{stok})
over the segments of the rod and the integral over infinity. Consequently with this remark it leads
to the following:
\be
\p \Sigma = \p \Sigma_{1} + \p \Sigma_{2} + \p \Sigma_{3} + \p \Sigma_{4} + \p \Sigma_{\infty},
\ee
we have denoted
\ben
\p \Sigma_{1} &=& \{ \rho = 0,~ z \in [- \infty,~ -ck^2] \}, \\ \nonumber
\p \Sigma_{2} &=& \{ \rho = 0,~ z \in [- ck^2,~ ck^2] \}, \\ \nonumber
\p \Sigma_{3} &=& \{ \rho = 0,~ z \in [ ck^2,~ k^2] \}, \\ \nonumber
\p \Sigma_{4} &=& \{ \rho = 0,~ z \in [ k^2,~ \infty ] \},\\ \nonumber
\p \Sigma_{\infty} &=& \{ (\rho,~z) \mid \sqrt{\rho^2 + z^2} \rightarrow \infty ~with~ z/ \sqrt{\rho^2 + z^2}~ finite \}.
\een
Having in mind relations (\ref{sl}) and (\ref{sm}) we can establish
that $\sigma(\varphi)_{,\la},
\sigma(\varphi)_{,\mu}$ and $e^{- \sigma(\varphi)}$ remain finite along the boundaries
$\p \Sigma^{(i)}$ for $i = 1,~ 2,~ 3,~4$ 
and they all
vanish along these parts 
of the boundary. 
Just, it remains to consider
the last part of the boundary $\p \Sigma_{\infty}$. In order to do so we introduce the coordinate
$(r, ~\theta)$ defined by $\rho = {r^2 \over 2} \sin 2 \theta$ and $z = {r^2 \over 2} \cos 2 \theta$.
Using Eqs.(\ref{m1}) and (\ref{m2}) in $(r,~ \theta)$
coordinates one reaches to the following
expressions for $\sigma(\varphi)_{,r}$ and $\sigma(\varphi)_{,\theta}$:
\ben \label{sl}
\sigma(\varphi)_{,r} &=&
{G_{AB} \over 2 }
\bigg[
{r \over \sin^2 2 \theta}~\varphi^{A}_{,r} \varphi^{B}_{,r} +
{ctg 4 \theta \over 2 r}~\varphi^{A}_{,\theta} \varphi^{B}_{,\theta} -
{ctg 4 \theta \over 2 \cos^2 2 \theta}~\varphi^{A}_{,r} \varphi^{B}_{,\theta} 
\bigg], \\ \label{sm} 
\sigma(\varphi)_{,\theta} &=&
{G_{AB} \over 4 }
\bigg[
- r^2~ ctg 2 \theta~ \varphi^{A}_{, r} \varphi^{B}_{, r} +
{1 \over \sin 2 \theta}~ \varphi^{A}_{,\theta} \varphi^{B}_{,\theta} +
4 r \cos^2 2 \theta ~ \varphi^{A}_{, r} \varphi^{B}_{,\theta}
\bigg].
\een
Hence, in terms of Eqs.(\ref{sl}) and (\ref{sm}) 
we arrive at the following relation:
\be
D_{\p \Sigma_{\infty}} = 
\lim_{r \rightarrow \infty}~
\int~d \theta~
\bigg(
e^{- \sigma(\varphi)}~ r^3 \sigma_{,r}~ \sin 2 \theta
- e^{- \sigma(\varphi)}~
r^2~\sigma_{, \theta}~\sin 2 \theta ~ctg 4 \theta
+ {\cal O}\bigg( {1 \over r^{n}} \bigg)
\bigg),
\ee
where $n \ge 2$.
Having in mind the asymptotical properties of the derivatives of scalar field $\varphi$
\be
\varphi^{A}_{, \theta} = {\cal O}\bigg( {1 \over r^{3/2}} \bigg),
\qquad
\varphi^{A}_{, r} = {\cal O}\bigg( {1 \over r^{5/2}} \bigg),
\ee
we conclude that
the above entire integral vanishes to the fact that 
$\lim_{r \rightarrow \infty} r^3~ \sigma_{,r} = 0$ and $\lim_{r \rightarrow \infty} r^2~ \sigma_{, \theta} = 0$.\\ 
Hence
${}^{(g)}\na^{i}  \sigma(\varphi) {}^{(g)}\na_{i} \sigma(\varphi)$ and 
${1 \over \rho} {}^{(g)}\na_{\rho}\sigma(\varphi) +
{}^{(g)}\na^{i} {}^{(g)}\na_{i}\sigma(\varphi) $ are equal to zero. 
It occurs that    
$\sigma(\varphi)$
is constant in the considered domain $\Sigma$, but using the fact that 
$\sigma(\varphi)$ tends to zero as $r \rightarrow \infty$ we get that $\sigma(\varphi) = 0$
which in turn implies that $\varphi$ is constant in the entire domain $\Sigma$. Just
from Eq.(\ref{ss}) one can deduced that $\sigma(vac)$ is the only solution of 
equations of motion. 
%%%%%%%%%%%%%%%%%%%%%%%%%%%%%%%%%%%%%%%%%%%%%%%%%%%%%%%%%%%%%%%%%%%%%%%%%%%%%%%%%%%%%%%%%%%%%%%%
In Ref.\cite{mor08}
uniqueness of the asymptotically flat, stationary five-dimensional black ring solution
being the solution of Einstein vacuum equations with regular
event horizon homeomorphic to $S^1 \times S^2$ and admitting three commutating
Killing vector fields (two spacelike and one timelike),
specified by mass and two angular momenta and the ratio of the radius of $S^2$
to $S^1$ was shown. 
Thus, we can assert the main conclusion of our work:\\
{\it Theorem}:\\
Let us consider a stationary axisymmetric solution to five-dimensional self-gravitating 
non-linear $\sigma$-models with an asymptotically timelike Killing vector field $k_{\mu}$
and two spacelike Killing vector fields $\phi_{\mu}$ and $\psi_{\mu}$.
The scalar field is invariant under the action of the Killing vector fields. Then, the only
asymptotically flat
black ring solution with regular rotating event horizon
is the five-dimensional PS vacuum black ring solution
with a constant mapping $\varphi$.

%%%%%%%%%%%%%%%%%%%%%%%%%%%%%%%%%%%%%%%%%%%%%%%%%%%%%%%%%%%%%%%%%%%%%%%
\begin{acknowledgments}
MR is grateful for hospitality of Institut des Hautes Etudes Scientifiques
where the part of the research was begun.  This
work was partially financed by the Polish budget funds in 2008 year as
the research project.
\end{acknowledgments}
%%%%%%%%%%%%%%%%%%%%%%%%%%%%%%%%%%%%%%%%%%%%%%%%%%%%%%%%%%%%%%%%%%%%%%%

%%%%%%%%%%%%%%%%%%%%%%%%%%%%%%%%%%%%%%%%%%%%%%%%%%%%%%%%%%%%%%%%%%%%%%%
%\begin{appendix}

%\section{Irred   } 
%\label{irtf}
%\end{appendix}
%%%%%%%%%%%%%%%%%%%%%%%%%%%%%%%%%%%%%%%%%%%%%%%%%%%%%%%%%%%%%%%%%%%%%%%%

\end{document}